\documentstyle[amssymb,cite]{elsart} 
\newcommand{\be}{\begin{equation}}   \newcommand{\ee}{\end{equation}} 
\newcommand{\ba}{\begin{eqnarray}}   \newcommand{\ea}{\end{eqnarray}} 
\newcommand{\ds}{\displaystyle} 
\newcommand{\mean}[1]{\left\langle #1 \right\rangle}

\newcommand{\ul}[1]{\underline{#1}} 
\newcommand{\non}{\nonumber \\} 
 
\newcommand{\eqn}[1]{Eq. (\ref{#1})} 
\newcommand{\Eqn}[1]{Eq. (\ref{#1})} 
\newcommand{\eqs}[2]{Eqs. (\ref{#1}), (\ref{#2})} 
 
\newcommand{\name}[1]{{\rm #1}}

\newcommand{\D}{\displaystyle} 
 
\begin{document} 
\begin{frontmatter} 
\title{ Phase transitions in social impact models of opinion formation } 
\author[pw,hu,mpi]{Janusz A. Ho\l yst\thanksref{emjh}}, 
\author[pw,mpi]{Krzysztof Kacperski\thanksref{emkk}}, 
\author[hu,bo]{Frank Schweitzer\thanksref{emfs}} 
\address[pw]{Faculty   of   Physics,   Warsaw University 
of Technology, Koszykowa 75, \\ \hspace*{0.15em} PL-00--662 
Warsaw, Poland } 
\address[hu]{Institute of Physics, Humboldt University, Unter den Linden 
6,\\ \hspace*{0.15em}  D-10099 Berlin, Germany} 
\address[mpi]{ Max Planck Institute for Physics of Complex Systems, 
N\"othnitzer Str. 38,\\ D-01187 Dresden, Germany } 
\address[bo]{GMD Institute for Autonomous intelligent Systems, Schloss 
  Birlinghoven,\\ \hspace*{0.15em} D-53754 Sankt Augustin, Germany} 
\thanks[emjh]{Corresponding author. Address: 
Faculty   of   Physics,   Warsaw University of Technology, 
Koszykowa 75, PL-00--662 Warsaw, Poland; tel.:+48 (22) 6607133, 
fax: +48 (22) 6282171, e-mail: jholyst@if.pw.edu.pl} 
\thanks[emkk]{e-mail: kacper@mpipks-dresden.mpg.de} 
\thanks[emfs]{e-mail: schweitzer@gmd.de} 
\date{15 March 2000} 
\begin{abstract} 
  We study phase transitions in models of opinion formation which are 
  based on the social impact theory. Two different models are discussed: 
  (i) a cellular--automata based model of a finite group with a strong 
  leader where persons can change their opinions but not their spatial 
  positions, and (ii) a model with persons treated as active Brownian 
  particles interacting via a communication field. In the first model, 
  two stable phases are possible: a cluster around the leader, and a 
  state of social unification. The transition into the second state 
  occurs for a large leader strength and/or for a high level of social 
  noise. In the second model, we find three stable phases, which 
  correspond either to a ``paramagnetic'' phase (for high noise and 
  strong diffusion), a ``ferromagnetic'' phase (for small nose and weak 
  diffusion), or a phase with spatially separated ``domains'' (for 
  intermediate conditions). 
\end{abstract} 
\end{frontmatter} 
\newpage 
 
\section{Introduction} 
During the last years there has been a great interest in 
applications of statistical physics in social science 
\cite{haken,complex}. Usually economical models are studied using 
the techniques of stochastic dynamics \cite{Stanley}, percolation 
theory \cite{Staufer} or the chaos paradigm \cite{chaos}. Another 
important subject of this kind is the process of {\it opinion 
formation} treated as a collective phenomenon. On the 
``macroscopic'' level it can be described using the master 
equation or Boltzmann-like equations for global variables 
\cite{weid,weid2,helb}, but microscopic models are constructed and 
investigated as well \cite{galam,bahr} using standard methods of 
statistical physics. A quantitative approach to the dynamics of 
opinion formation is related to the concept of \emph{social 
impact} \cite{lata,val,impact,lew,nsl,plew,kohring}, which enables 
to apply the methods similar to the cellular automata approach 
\cite{galam,ca}. 
 
The aim of this work is to study various kinds of phase 
transitions in two models  based on the social impact theory. In 
Sec. 2 we consider phase transitions in a social impact model that 
can occur in a finite group in the presence of a {\it strong 
individual} (a leader)\cite{jsp96,berl,PhysA}.  As two special 
cases, we discuss a purely deterministic limit and a noisy model. 
Sec. 3 is devoted to an extension of social impact models to 
include phenomena of migration, memory effects and a finite 
velocity of information exchange. Here the concepts of {\it active 
  Brownian particles} 
\cite{lsg-mieth-rose-malch-95,lsg-schw-mieth-97,schw-agent-97} and the 
communication field \cite{SchwHol} will be applied. 
 
\section{Phase transitions in the presence of a strong leader} 
\subsection{The model} 
Our system consists of $N$ individuals (members of a social group); we 
assume that each of them can share one of two opposite opinions on a 
certain subject, denoted as $\sigma_{i} = \pm 1, \: i = 1,2,...N$. 
Individuals can influence each other, and each of them is characterised 
by the parameter $s_{i}>0$ which describes the strength of his/her 
influence.  Every pair of individuals $(i,j)$ is ascribed a distance 
$d_{ij}$ in a social space.  The changes of opinion are determined by the 
{\it social impact} exerted on every individual: 
\be I_{i}=  - s_i \beta - \sigma_i h - 
\sum_{j=1, j\neq i}^{N} \frac{s_{j}\sigma_{i}\sigma_{j}} 
{g(d_{ij})}, 
\label{e3} \ee 
where $g(x) $ is an increasing function of social distance. $ \beta$ is a 
so--called self--support parameter reflecting the inclination of an 
individual to maintain his/her current opinion. $h$ is an additional 
(external) influence which may be regarded as a global preference towards 
one of the opinions stimulated by mass--media, government policy, etc. 
 
Opinions of individuals may change simultaneously (synchronous dynamics) 
in discrete time steps according to the rule: 
\be \sigma_{i}(t+1) = \left\{ \begin{array}{rrr} \sigma_{i}(t) & 
\mbox{with probability} & 
\frac{\ds\exp(-I_{i}/T)}{\ds\exp(-I_{i}/T)+ \ds\exp(I_{i}/T)} 
\vspace{2mm} 
\\ -\sigma_{i}(t) &  \mbox{with probability} & \frac{ 
\ds\exp(I_{i}/T)} {\ds\exp(-I_{i}/T)+ \ds\exp(I_{i}/T)} 
\end{array} \right. \label{e14}. \ee 
\Eqn{e14} is analogous to the Glauber dynamics with 
$-I_{i}\,\sigma_{i}$ corresponding to the local field. The 
parameter $T$ may be interpreted as a ``social temperature'' 
describing a degree of randomness in the behaviour of individuals, 
but also their average volatility (cf \cite{bahr}). The impact 
$I_i$ is a ``deterministic'' force inclining the individual $i$ to 
change his/her opinion if $I_i>0$, or to keep it otherwise. The 
model is a particular case of the system considered in \cite{lew}. 
 
We assume that our social space is a 2D disc of radius $R \gg 1$, 
with the individuals located on the nodes of a quadratic grid. The 
distance between nearest neighbours equals 1, while the geometric 
distance models the social immediacy. The strength parameters $ 
s_i $ of the individuals are positive random numbers with 
probability distribution $q(s)$ and the mean value $ 
\overline{s}$. In the centre of the disc there is a {\it 
  strong individual} (who we will call the ``leader''); his/her strength 
$s_{L}$ is much larger than that of all the others $(s_{L} \gg s_i)$.

\subsection{Deterministic limit} 
\label{deter} 
Let us first recall the properties of the system without noise, i.e. at 
$T=0$ \cite{jsp96,berl}. Then, the dynamical rule (\ref{e14}) becomes 
strictly deterministic: 
\begin{equation}\label{new1} 
  \sigma_i(t+1)= -\mbox{sign}(I_i \sigma_i). 
\end{equation} 
Considering the possible stationary states we find either the trivial 
unification (with equal opinion $\pm 1$ for each individual) or, due to 
the symmetry, a circular cluster of individuals who share the opinion of 
the leader. This cluster is surrounded by a ring of their opponents (the 
majority).  These states remain stationary also for a small self-support 
parameter $\beta$; for sufficiently large $\beta$ any configuration may 
remain ``frozen''. 
 
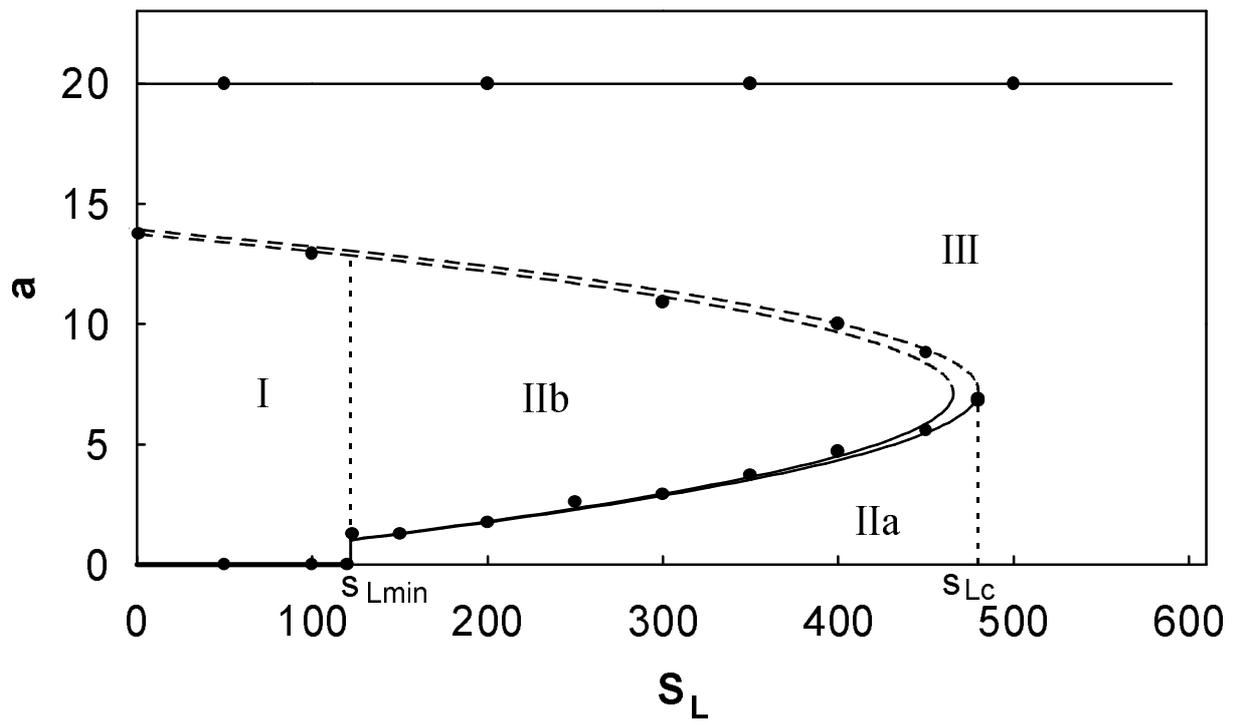
\begin{figure}[h] 
\caption{Cluster's radius $a$ vs. leader's strength $s_{L}$ -- 
phase diagram for circular social space. Interactions proportional 
to inverse of mutual distance ($I\propto 1/r$). Lines correspond 
to analytical results, points to computer simulations.} 
\label{a(sL)} 
\end{figure}

Using the approximation of continuous distribution of individuals (i.e. 
replacing the sum in (\ref{e3}) by an integral) one can calculate the 
size of the cluster, i.e. its radius $a$ as a function of the other 
parameters. In the case of $ g(r)=r $ and $\overline{s}=1$ we get from 
the limiting condition for the stationarity $I=0$ at the border of the 
cluster: 
\be a \approx \frac{1}{16} \left[ 2\pi R-\sqrt{\pi}\pm\beta-h 
  \pm \sqrt{(2\pi R-\sqrt{\pi}\pm\beta-h)^2-32s_L}\: \right]. \label{e8} 
\ee 
This is an approximate solution valid for $ a\ll R $, but it 
captures all the qualitative features of the exact one which can 
be obtained by solving a transcendent equation (cf. 
Fig.~\ref{a(sL)}). Here and in the next section we assume that the 
leader's opinion is $\sigma_L = +1$, but the analysis is also 
valid for the opposite case if $h \rightarrow -h$. 
 
The branch with the ``$ - $'' sign in front of the square root in 
\eqn{e8} corresponds to the stable cluster.  The one with ``+'' 
corresponds to the unstable solution which separates the basins of 
attraction of the stable cluster and unification (cf. 
Fig.~\ref{a(sL)}).  Owing to the two possible signs at the 
self--support parameter $\beta $ in (\ref{e8}), the stable and 
unstable solutions are split and form in fact two bands.  The 
states within the bands are ``frozen'' due to the self--support 
which may be regarded as an analogy of the dry friction in 
mechanical systems. This way also the unstable clusters can be 
observed for $\beta >0$ and appropriately chosen initial 
conditions. 
 
According to  Eq. (\ref{e8})  real solutions corresponding to 
clusters exist provided \be  (2\pi R-\sqrt{\pi}\pm\beta-h)^2-32s_L 
\geqslant 0. \label{pekc} \ee 
Otherwise the general acceptance of the leader's opinion (unification) is 
the only stable state. When, having a stable cluster, the condition 
(\ref{pekc}) is violated by changing a parameter e.g. $s_L$ or $h$, one 
can observe a discontinuous phase transition: {\it cluster} $\rightarrow$ 
{\it unification}. 
 
If, on the other hand, the leader's strength is too weak, it may be 
impossible for him/her not only to form a cluster but also to maintain 
his/her own opinion.  The limiting condition for the minimal leader's 
strength $s_{Lmin}$ to resist against the persuasive impact of the 
majority can be calculated from the limiting condition $I_L=0$ ($I_L$ - 
the impact exerted on the leader): 
\be s_{Lmin}=\frac{1}{\beta}(2\pi R-\sqrt{\pi}-h).\label{e10}\ee 

Fig.~\ref{a(sL)} shows a phase--diagram $ s_{L}\mbox{-}a $ for 
$h=0$. All the plots are made for a space of radius $R = 20$ (1257 
individuals) and $\beta = 1$ unless stated otherwise. Points in 
Fig.~\ref{a(sL)} are obtained by numerical simulations of 
(\ref{new1}) while the curves are solutions of a transcendent 
equation following from the stationary condition  $I(a)=0$. Solid 
lines represent stable fixed points -- attractors (they correspond 
to the solution (\ref{e8}) with ``$ - $'' sign before the square 
root); dashed lines represent unstable repellers (corresponding to 
``+'' in (\ref{e8})). 
 
We find two kinds of attractors: (i) unification ($ a = R $ when the 
leader's opinion wins, $ a = 0 $ when it ceases to exist) and (ii) a 
stable cluster resulting from a solution of (\ref{e8}). In the $ 
s_{L}\mbox{-}a \:$ space one can distinguish between three basins of 
attraction.  Starting from a state in the area denoted as $I$, the time 
evolution leads to unification with $a = 0$. The stable cluster attractor 
divides its basin of attraction into the areas $IIa$ and $IIb$. All 
states from $III$ will evolve to unification with $a=20$. Owing to the 
two possible signs of self--support parameter $\beta $ in (\ref{e8}), the 
attractor and repeller are split. The space between their two parts 
enclose the ``frozen'' states that do not change in the course of time. 
These states correspond to local equilibria of the system dynamics 
similar to spin glass states. Thus, as a result of self--support, even 
repeller states can be stabilized. As one can see, the results of 
computer simulations fit the calculated curves very well. 
 
Taking into account the conditions (\ref{pekc}), (\ref{e10}) and 
the two possible opinions of the leader  one can draw a 
phase-diagram $h-s_L$ distinguishing the regions where different 
system states are possible \cite{jsp96,PhysA}. Apparently, the 
system shows {\it multistability} in a certain range of $s_L$ and 
$h$. It depends on the history which of the states is realized, so 
we can observe a {\it hysteresis} phenomenon \cite{jsp96,PhysA}. 
Moving in the parameter space $s_L - h$, while starting from 
different configurations one can have many possible scenarios of 
phase transitions \cite{PhysA}. 
 
\subsection{Effects of social temperature} 
It is obvious that the behaviour of an individual in a group 
depends not only on the influence of others. There are many more 
factors, both internal (personal) and external, that induce 
opinion formation and should be modeled somehow. In our model, we 
do this by means of a noisy dynamics, i.e. we use the equation 
(\ref{e14}) with the parameter $T > 0$.  In the presence of noise, 
the marginal stability of unstable clusters due to the 
self-support is suppressed and they are no longer the stationary 
states of the system. The borders of the stable clusters become 
diluted, i.e. individuals of both opinions appear all over the 
group.  Our simulations \cite{jsp96,PhysA} prove that the presence 
of noise can induce the transition from the configuration with a 
cluster around the leader to the unification of opinions in the 
whole group. There is a well defined temperature $T_c$ that 
separates these two phases. To estimate the dependence of $T_c$ on 
other system parameters analytically, one can use a mean field 
approach, like  methods developed in \cite{jsp96,PhysA}. The two 
limiting cases of such an approach correspond to low temperature 
and high temperature approximations and are discussed in the 
following. 

\subsection{Low--temperature mean--field approximation} 
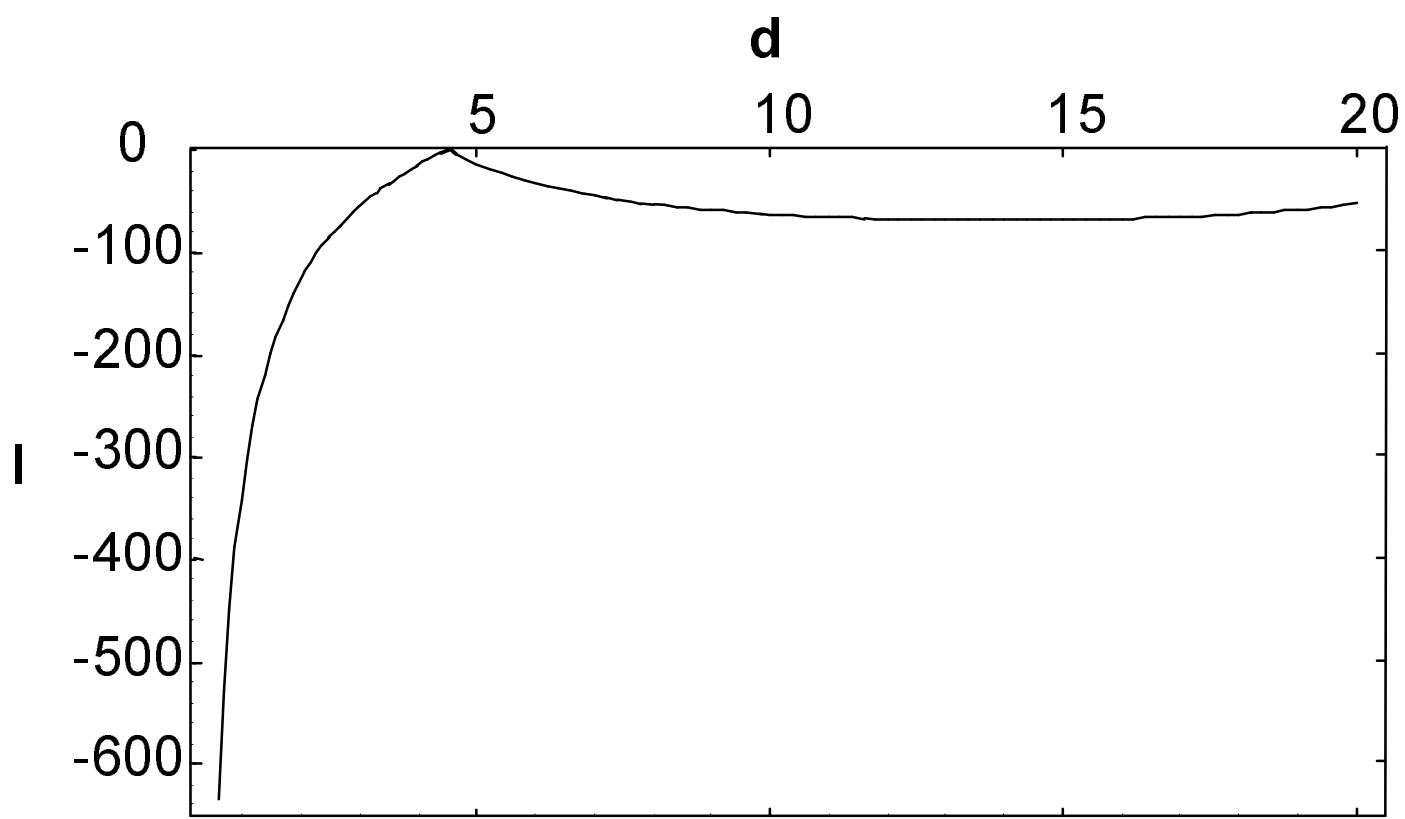
\begin{figure}[h] 
\caption{Social impact $I$ as a function of distance $d$ to the 
leader. Leader's strength $s_{L} = 400$.} 
\label{I(d)} 
\end{figure}

For {\it low temperatures} $T$, i.e. for a small noise level, the cluster 
of leaders followers is only slightly diluted and its {\it effective 
  radius} $a(T)$ can be treated as an order parameter.  One can then 
calculate the impact $I(d)$ acting on the group member inside ($d<a$) and 
outside ($d>a$) the cluster respectively \cite{jsp96}: 
\be I_{i}(d)=-\frac{s_{L}}{d}-8a\,E\!\left(\frac{d}{a}\,, 
\frac{\pi}{2}\right)+4R\, E\!\left(\frac{d}{R}\,, 
\frac{\pi}{2}\right)+2\sqrt{\pi}-\beta, \label{e21} \ee 
\be I_{o}(d)=\frac{s_{L}}{d}+8a\,E\!\left(\frac{d}{a}\,, 
\arcsin\frac{a}{d}\right)-4R\, 
E\!\left(\frac{d}{R}\,,\frac{\pi}{2} \right)+2\sqrt{\pi}-\beta, 
\label{e22} \ee 
where $E(k,\varphi)=\int_0^\varphi (1-k^2 
\sin^2\alpha)^{1/2}d\alpha$ is the elliptic integral of the second 
kind. Both functions are plotted in Fig.~\ref{I(d)} for $s_{L} = 
400$. The system remains in equilibrium, therefore the impact on 
every individual is negative (nobody changes his/her opinion). It 
approaches zero at the border of the cluster which means that 
individuals located in the neighbourhood of that border are most 
sensitive to thermal fluctuations. We can however observe a 
significant {\it asymmetry} of the impact. It is considerably 
stronger inside the cluster. Individuals near the leader are 
deeper confirmed in their opinion, so they are also more resistant 
against noise in dynamics. When we increase the temperature 
starting from $T\simeq 0$, random opinion changes begin. Primarily 
it concerns those near the border (the weakest impact). As a 
result individuals with adverse opinions appear both inside and 
outside the cluster. They are more numerous outside because of the 
weaker impact (cf. Fig.~\ref{I(d)}). 
 
Effectively, we observe the growth of a minority group. This causes the 
supportive impact outside cluster to become still weaker and the majority 
to become more sensitive to random changes. It is a kind of positive 
feedback. At certain value of temperature the process becomes an 
avalanche, and the former majority disappears. Thus, noise induces a jump 
from one attractor (cluster) to another (unification). Such a transition 
is possible for every non-zero temperature, but its probability remains 
negligible until the noise level exceeds a certain critical value that 
corresponds to our critical temperature $T_c$. 
 
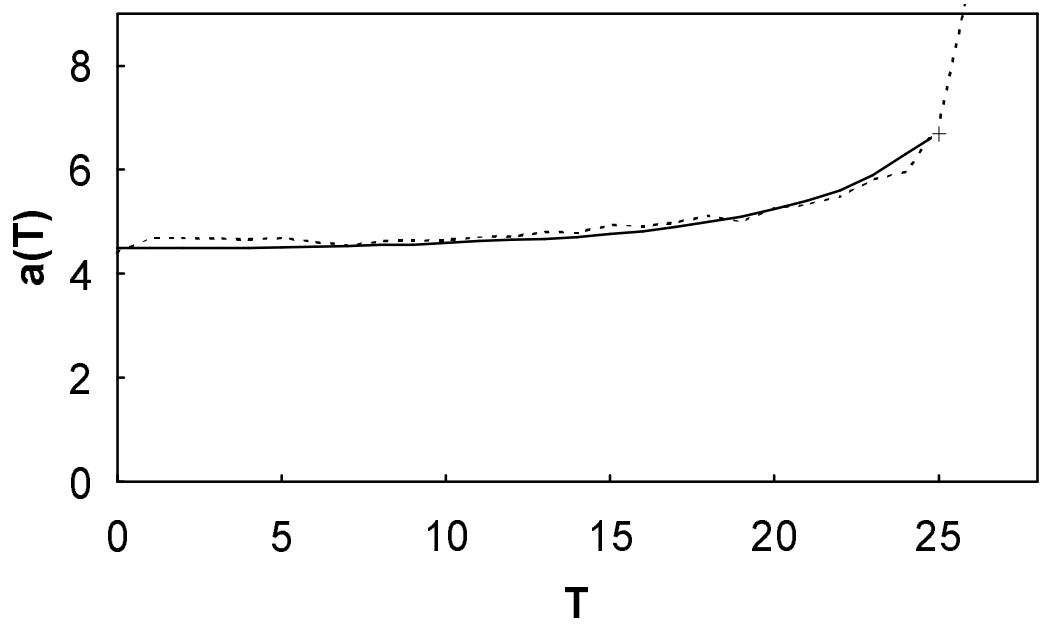
\begin{figure}[h] 
\caption{ Mean cluster radius $a$ vs. temperature $T$; $s_{L} = 
400$. Results of calculation (solid) and computer simulation 
(dotted).} 
\label{a(T)} 
\end{figure}

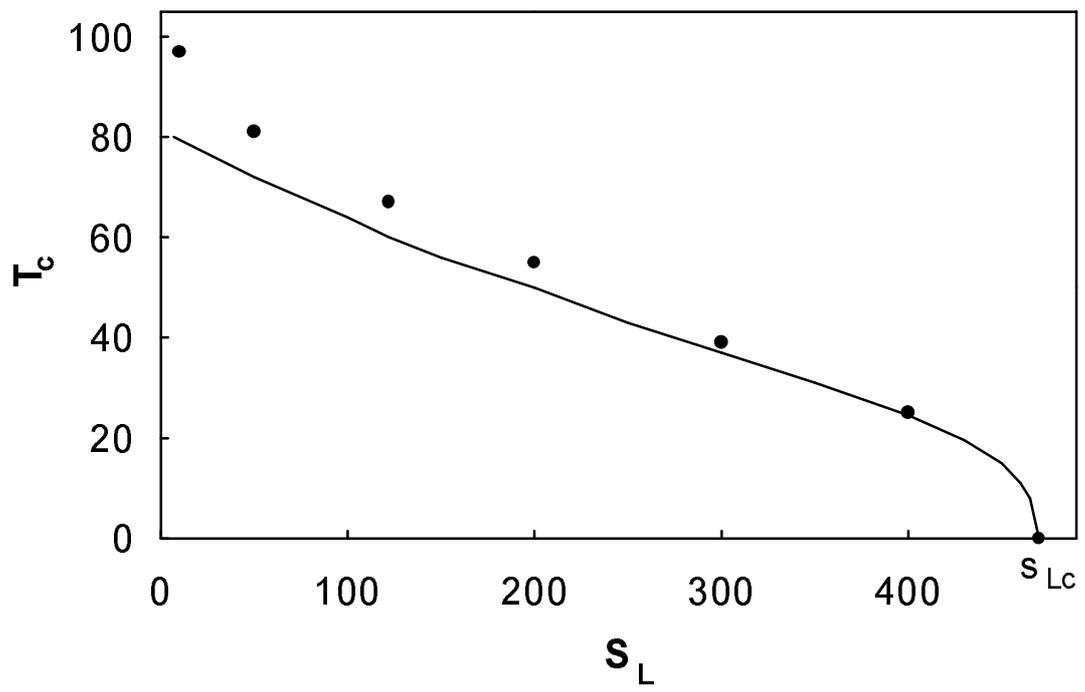
\begin{figure}[h] 
\caption{Critical temperature $T_{c}$ (above which no stable 
cluster exists) vs. leader's strength $s_{L}$. Leader's opinion 
fixed (independent of the group). Line -- calculations 
(Eq.~(\ref{e25})), points -- simulations. } 
\label{Tc(sL)} 
\end{figure}

Using Eq. (\ref{e14}) and taking into account Eqs. (\ref{e21}) and 
(\ref{e22}) we can compute the probability $ 
\mbox{Pr}(\sigma=1)(r)$ that an individual at the distance $r$ 
from the leader, shares opinion +1, which is assumed as the 
opinion of the leader.  Then, the mean number of all individuals 
with opinion $+1$ may be calculated by integrating this 
probability multiplied by the surface density (equaling 1) over 
the whole space: \be \overline{n(\sigma=1)(T)}=\int_{0}^{R}\! 
\mbox{Pr}(\sigma=1)(r)\, 2\pi r\,dr. \label{e24} \ee 
This number equals the effective area of the circular cluster, so 
its radius is 
\be \overline{a(T)}=\sqrt{\frac{\overline{n(\sigma=1)(T)}}{\pi}}. 
\label{e25} \ee 
The  \eqn{e25} is a rather involved transcendent equation for 
$a(T)$ (it appears on the right hand side in $I_{i}(r)$ and 
$I_{o}(r))$. For low temperatures $T$ it has three solutions 
$a(T)$ corresponding to a stable cluster, an unstable cluster and 
a social homogeneous state.  The numerical solution for the radius 
of stable cluster is presented in Fig. ~\ref{a(T)} together with 
results of computer simulations. One should point out that the 
radius of the cluster $a$ is an increasing function of the 
temperature $T$ for the reasons discussed above.  At some critical 
temperature, a pair of solutions corresponding to the stable and 
the unstable cluster coincide \cite{jsp96,PhysA}. Above this 
temperature, there exists {\it only} the solution corresponding to 
the social homogeneous state. Fig. ~\ref{Tc(sL)} shows the plot of 
the critical temperature $T_c$ obtained from (\ref{e25}) as the 
function of the leader strength $s_L$ together with results of 
computer simulations. 
 
\subsection{High--temperature mean--field approximation} 
For high temperatures or small values of the leader's strength $s_L$, the 
cluster around the leader is very diluted and it is more appropriate to 
assume that there is a {\it spatially homogeneous mixture} of leaders 
followers and opponents, instead of a {\it localized cluster} with a 
radius $a(T)$.  It follows that at each site there is {\it the same} 
probability $0< p(T) < 1$ to find an individual sharing the leaders 
opinion, and $p(T)$ plays the role of order parameter. Neglecting the 
self--support $(\beta=0)$ one can write the social impact acting on a 
opponent of the leader at place $x$ as \cite{PhysA}: 
\be I(x)=\frac{s_L}{g(x)}+(2 p-1) \rho  \overline{s} J_D(x) + h 
\label{impr1} \ee 
$ J_D(x) = \int_{D_R} 1/g(|{\bf r}-{\bf x}|) d^2 {\bf r}$ is a 
function which depends only on the size of the group and the type 
of interactions. After a short algebra one gets the following 
equation for the probability $p(T)$\cite{PhysA}: 
\be p = 
\frac{1}{\pi R^2 \rho} \int_{0}^{R}\!\rho \: \mbox{Pr}(r)\, 2\pi 
r\,dr = \frac{1}{ R^2} \int_{0}^{R} 
\frac{\exp{[I(r,p)/T]}}{\cosh{[I(r,p)/T]}} r dr \equiv f(p), 
\label{e23b} \ee 
where $I(x,p)$ is given by (\ref{impr1}). Similar to equation (\ref{e25}) 
obtained for low temperatures, there are three solutions of \eqn{e23b}: 
the smallest one corresponds to the stable cluster around the leader, the 
middle one to the unstable cluster which, in fact, is not observed, and 
the largest one to the unification. The size of the stable cluster grows 
with increasing temperature up to a critical value $T_{c}$ when it 
coincides with the unstable solution. At this temperature, a transition 
from a stable cluster to unification occurs \cite{PhysA}. For $T > 
T_{c}$, unification is the only solution, but it is no longer a perfect 
unification because due to the noise individuals of the opposite opinion 
appear. When the temperature increases further, $p(T)$ tends to $1/2$ 
which means that the dynamics is random and both opinions appear with 
equal probability. 
 
\section{Modelling  opinion dynamics by means of active Brownian 
  particles} 
\subsection{The model} 
There are several basic disadvantages of the  model considered in 
the previous chapter. In particular, it assumes, that the impact 
on an individual is updated with infinite velocity, and no memory 
effects are considered. Further, there is no migration of the 
individuals, and any ``spatial'' distribution of opinions refers 
to a ``social'', but not to the physical space. 
 
An alternative approach \cite{SchwHol} to the social impact model of 
collective opinion formation, which tries to include these features is 
based on {\em active Brownian particles} \cite{lsg-mieth-rose-malch-95, 
  lsg-schw-mieth-97, schw-agent-97, fs-eb-tilch-98-let, eb-fs-tilch-98}, 
which interact via a {\it communication field}.  This scalar field 
considers the spatial distribution of the individual opinions, further, 
it has a certain life time, reflecting a collective memory effect and it 
can spread out in the community, modeling the transfer of information. 
 
The spatio-temporal change of the communication field is given by 
the following equation: 
\begin{equation} 
\label{hrt} \frac{\partial}{\partial t} h_{\sigma}({\bold r},t) = 
\sum_{i=1}^{N}s_{i}\delta_{\sigma,\sigma_{i}}\; \delta({\bold 
r}-{\bold r}_{i})\ - \gamma h_{\sigma}({\bold r},t) + D_{h} \Delta 
h_{\sigma}({\bold r},t). 
\end{equation} 
Every individual contributes permanently to the field $h_{\sigma}({\bold 
  r},t)$ with its opinion $\sigma_{i}$ and with its personal strength 
$s_{i}$ at its current spatial location ${\bold r}_{i}$.  Here, 
$\delta_{\sigma,\sigma_{i}}$ is the \name{Kronecker} Delta, 
$\delta({\bold r}-{\bold r}_{i})$ denotes \name{Dirac's} Delta 
function used for continuous variables, $N$ is the number of 
individuals. The information generated by the individuals has a 
certain average life time $1/\gamma$, further it can spread 
throughout the system by a diffusion-like process, where $D_{h}$ 
represents the diffusion constant for information exchange. If two 
different opinions are taken into account, the communication field 
should also consist of two components, $\sigma=\{-1,+1\}$, each 
representing one opinion. 
 
In this model, the scalar {\em spatio-temporal communication 
field} $h_{\sigma}({\bold r},t)$ \cite{SchwHol}, plays in part the 
role of social impact $I_i$ used in \cite{jsp96,PhysA}.  Instead 
of a social impact, the communication field $h_{\sigma}({\bold 
r},t)$ influences the individual $i$ as follows: At a certain 
location ${\bold r}_{i}$, the individual with opinion 
$\sigma_{i}=+1$ is affected by two kinds of information: the 
information resulting from individuals who share his/her opinion, 
$h_{\sigma=+1}({\bold r}_{i},t)$, and the information resulting 
from the opponents $h_{\sigma=-1}({\bold r}_{i},t)$.  Dependent on 
the {\em local} information, the individual reacts in two ways: 
(i) it can \emph{change its opinion}, (ii) it can \emph{migrate} 
towards locations which provide a larger support of its current 
opinion. These opportunities are specified in the following. 
 
We assume that the probability $p_{i}(\sigma_{i},t)$ to find the 
individual $i$ with the opinion $\sigma_{i}$ changes in the course 
of time due to the  master equation (the dynamics is continuous in 
time): 
\begin{equation} 
\frac{d}{d t}p_{i}(\sigma_{i},t)=\sum_{\sigma_{i}'} 
w(\sigma_{i}|\sigma_{i}') p_{i}(\sigma_{i}',t) - 
p_{i}(\sigma_{i},t)\sum_{\sigma_{i}'} w(\sigma_{i}'|\sigma_{i}). 
\label{p-eins} 
\end{equation} 
where rates of transition probability are  described in a similar 
way to Eq. (\ref{e14})

\begin{equation} 
w(\sigma_{i}'|\sigma_{i})= \eta  \exp\{[h_{\sigma'}({\bold 
r}_{i},t)-h_{\sigma}({\bold r}_{i},t)]/T\}\mbox{\hspace{0.5cm}for 
\hspace{0.5cm}}\sigma\neq\sigma' \label{wh} 
\end{equation} 
and $w(\sigma_{i}|\sigma_{i})= 0$. The movement of the individual 
located at space coordinate ${\bold 
  r}_{i}$ is described by the following overdamped Langevin equation: 
\begin{equation} 
\frac{d{\bold r}_i}{dt}=\alpha_{i} \left. \frac{\partial 
h_e({\bold r},t)}{\partial r}\right|_{{\bold r}_i} + 
\sqrt{2\,D_{n}}\;\xi_i(t). 
\label{langev-red} 
\end{equation} 
In the last term of \eqn{langev-red} $D_{n}$ means the spatial 
diffusion coefficient of the individuals. The random influences on 
the movement are modeled by a stochastic force with a 
$\delta$-correlated time dependence, i.e. $\xi(t)$ is white noise 
with $\mean{\xi_{i}(t)\,\xi_{j}(t')}=\delta_{ij}\,\delta(t-t')$. 
The term $h_e({\bold r},t)$ in \eqn{langev-red} means an {\em 
effective} communication field which results from 
$h_{\sigma}({\bold r},t)$ as a certain function of both 
components, $h_{\pm 1}({\bold r},t)$ \cite{SchwHol}. Parameters 
$\alpha_{i}$ are  individual response parameters. In the 
following, we will assume $\alpha_{i}=\alpha$ and 
$h_{e}=h_{\sigma}$. 
 
\subsection{Critical conditions for spatial opinion separation} 
The spatio-temporal density of individuals with opinion $\sigma$ can be 
obtained as follows: 
\begin{equation} 
n_{\sigma}({\bold r},t)=\int \sum_{i=1}^N 
\delta_{\sigma,\sigma_{i}} \delta({\bold r}-{\bold r}_{i}) 
P(\sigma_{1},{\bold r}_{1}...,\sigma_{N},{\bold r}_{N},t) d{\bold 
r}_1 ... d{\bold r}_{N} \label{dens} 
\end{equation} 
$P(\ul{\sigma},\ul{r},t)=P(\sigma_{1},{\bold r}_{1},...,\sigma_{N}, 
{\bold r}_{N},t)$ is the canonical $N$-particle distribution function 
which gives the probability to find the $N$ individuals with the opinions 
$\sigma_{1},...,\sigma_{N}$ in the vicinity of ${\bold r}_1,....,{\bold 
  r}_N$ on the surface $A$ at time $t$. The evolution of 
$P(\ul{\sigma},\ul{r},t)$ can be described by a master equation 
\cite{SchwHol} which considers both \eqs{wh}{langev-red}. 
Neglecting higher order correlations, one obtains from \eqn{dens} 
the following reaction-diffusion equation for $n_{\sigma}({\bold 
r},t)$ \cite{lsg-schw-mieth-97, SchwHol}: 
\begin{eqnarray} 
\label{fpe-t} \frac{\partial}{\partial t}\,n_{\sigma}({\bold r},t) 
= & - & \nabla\,\Big[n_{\sigma}({\bold r},t)\, \alpha \, \nabla 
h_{\sigma}({\bold r},t)\Big] + D_n\,\Delta n_{\sigma}({\bold r},t) 
\nonumber \\ & - & \sum_{\ul{\sigma'} \neq \ul{\sigma}} \Big[ 
w(\sigma'|\sigma)\,n_{\sigma}({\bold r},t) + 
w(\sigma|\sigma')\,n_{\sigma'}({\bold r},t)\Big] \non 
\end{eqnarray} 
with the transition rates given by eq. (\ref{wh}).  Eq. 
\ref{fpe-t} together with Eq. \ref{hrt} form a set of four 
equations describing our system completely. 
 
Now, let us assume that the spatio-temporal communication field 
\emph{relaxes faster} than the related distribution of individuals 
to a quasi-stationary equilibrium. The field $h_{\sigma}({\bold 
r},t)$ should still depend on time and space coordinates, but, due 
to the fast relaxation, there is a fixed relation to the 
spatio-temporal distribution of individuals. Further, we neglect 
the independent diffusion of information, assuming that the 
spreading of opinions is due to the migration of the individuals. 
Using $\dot{h}_{\sigma}({\bold r},t)=0$, $s_{i}=s$ and $D_{h}=0$ 
we get: 
\begin{equation} 
  \label{fixed} 
  h_{\sigma}({\bold r},t)=\frac{s}{\gamma}\,n_{\sigma}({\bold r},t) 
\end{equation} 
Inserting \eqn{fixed} into \eqn{fpe-t} we reduce the 
set of coupled equations to two equations. 
 
The homogeneous solution for $n_{\sigma}({\bold r},t)$ is given by the 
mean densities: 
\begin{equation} 
  \label{hom} 
\bar{n}_{\sigma}=\frac{\bar{n}}{2}\mbox{\hspace{2cm}where\hspace{2cm}} 
\bar{n}=\frac{N}{A} 
\end{equation} 
Under certain conditions however, the homogeneous state becomes 
unstable and a spatial separation of opinions occurs. In order to 
investigate these critical conditions, we allow small fluctuations 
$\delta n_{\sigma} \sim \exp \left( \lambda t+i {\bold k} {\bold 
r} \right)$ around the homogeneous state $\bar{n}_{\sigma}$ and 
perform linear stability analysis \cite{SchwHol}. The resulting 
dispersion relations read: 
\begin{equation} 
\label{dispers} 
\begin{array}{l} 
\lambda_{1}({\bold k})= - k^{2}\,C+2B\;;\;\; \lambda_{2}({\bold 
k})= - k^{2}\,C \\ B=\frac{\D \eta \,s\,\bar{n}}{\D \gamma T}-\eta 
\;;\;\; C=D_{n} - \frac{\D \alpha s\,\bar{n}}{\D 2 \gamma} 
\end{array} 
\end{equation} 
It follows that stability conditions of the homogeneous state, 
\mbox{$n_{\sigma}({\bold r},t)= \bar{n}/2$}, can be expressed as: 
\begin{equation} 
  \label{crit-temp} 
  T > T^{c}_{1}= \frac{s\,\bar{n}}{\gamma}, 
\mbox{\hspace{3cm}} D > D^{c}_{n}=\frac{\alpha}{2}\; 
\frac{s\,\bar{n}}{\gamma} 
\end{equation} 
If the above conditions are not fulfilled, the homogeneous state 
that corresponds to {\it paramagnetic phase} is unstable (i) 
against the formation of spatial ``domains'' where one of opinions 
 $\sigma=\pm1$ {\it locally dominates}, or  (ii) against the formation of a 
{\it ferromagnetic} state where the {\it total numbers} of people 
sharing both opinions are not equal. 
 
Case (i) can occur only for a systems whose linear dimensions are large 
enough, so that large--scale fluctuations with small wave numbers can 
destroy the homogeneous state \cite{SchwHol}.  In case (ii), each 
subpopulation can exist either as a \emph{majority} or as a 
\emph{minority} within the community.  Which of these two possible 
situations is realized, depends in a deterministic approach on the 
initial fraction of the subpopulation.  Breaking the symmetry between the 
two opinions due to {\it external influences} (support) for one of the 
opinions would increase the region of initial conditions which lead to a 
majority status. Above a critical value of such a support, the 
possibility of a minority status completely vanishes and the supported 
subpopulation will grow towards a majority, regardless of its initial 
population size, with no chance for the opposite opinion to be 
established \cite{SchwHol}. 
 
\section{Conclusions} 
This work discusses the possibilities of phase transitions in 
models of opinion formation which are based on the social impact 
theory (two opinions case). In the presence of a strong leader 
situated in the centre of a finite group, a transition can take 
place from a state with a cluster around the leader to a state of 
uniform opinion distribution where virtually all members of the 
group share the leaders's opinion. The transition occurs if a 
leader's strength exceeds a well defined critical value or if the 
noise level (``social temperature'') is high enough.  The weaker 
the leader's strength is, the larger noise is needed. The value of 
the critical temperature can be calculated using mean field 
methods where either the existence of an effective value of the 
cluster radius (low temperature method) or a spatially homogeneous 
mixture of both opinions (high temperature method) is assumed. 
Numerical simulations confirm the analytic results. 
 
The extension of the social impact model is based on the concept of 
active Brownian particles which communicate via a scalar, multi-component 
communication field. This allows us to take into account effects of 
spatial migration (drift and diffusion), a finite velocity of information 
exchange and memory effects. The reaction-diffusion equation for the 
density of individuals with a certain opinion is obtained. In this model, 
the transition can take place between the ``paramagnetic'' phase, where 
the probability to find any of opposite opinions is $1/2$ at each place 
(a high temperature and a high diffusion phase), the ``ferromagnetic'' 
phase with a global majority of one opinion (a low temperature and a low 
diffusion phase) and a phase with spatially separated ``domains'' with a 
local majority of one opinion (an intermediate phase). 
 
\section*{Acknowledgements} 
The work of one of us (JAH) has been financed in part by SFB 555 
{\it Komplexe Nichtlineare Prozesse}. 
 
\newpage 

\newpage 
\section*{Figure Captions} 
 
\clearpage 
\end{document}